\documentclass[12pt,a4paper,DIV12]{scrartcl}
\usepackage[utf8x]{inputenc}
\usepackage[T1]{fontenc}
\usepackage{lmodern}
\usepackage[british]{babel}
\usepackage{graphicx}
\usepackage{grffile}
\usepackage{hyperref}
\usepackage{color}
\usepackage{subfigure}
\usepackage{amsmath}
\usepackage{amssymb}
\usepackage{amsfonts}
\newcommand{\I}{\ensuremath{\mathrm{i}}}
\newcommand{\E}{\ensuremath{\mathrm{e}}}
\newcommand*{\email}[1]{\href{mailto:#1}{#1}} 
\newcommand{\arxiv}[2]{[arXiv:\,\href{http://arxiv.org/abs/#1}{\texttt{#1}} [\texttt{#2}]]}
\newcommand{\arxivold}[1]{[arXiv:\,\href{http://arxiv.org/abs/#1}{\texttt{#1}}\,]}
\newcommand{\mps}{\ensuremath{m_{\text{PS}}}}
\newcommand{\mv}{\ensuremath{m_{\text{V}}}}
\newcommand{\ms}{\ensuremath{m_{\text{S}}}}
\newcommand{\mpcac}{\ensuremath{m_{\text{\tiny PCAC}}}}
\newcommand{\mpv}{\ensuremath{m_{\text{PV}}}}
\newcommand{\msh}{\ensuremath{m_{1/2}}}
\newcommand{\mgb}{\ensuremath{m_{0^{++}}}}

\title{%
   {\vspace{-20mm}\normalsize
    \hfill\parbox[b][30mm][t]{35mm}{\textmd{MS-TP-17-28\\DESY 17-184}}}\\[-18mm]
Low energy properties of SU(2) gauge theory with \boldmath{$N_f = 3/2$} flavours
of adjoint fermions
\vspace*{2mm}}
\author{%
Georg Bergner\\
\textit{\large University of Jena, Institute for Theoretical Physics}\\
\textit{\large Max-Wien-Platz 1, D-07743 Jena, Germany}\\
\textit{\large E-mail: \email{georg.bergner@uni-jena.de}}\\[5mm]
Pietro Giudice, Gernot M\"unster, Philipp Scior\\
\textit{\large University of M\"unster, Institute for Theoretical Physics}\\
\textit{\large Wilhelm-Klemm-Str.~9, D-48149 M\"unster, Germany}\\
\textit{\large E-mail: \email{\{p.giudice,munsteg,scior\}@uni-muenster.de}}\\[5mm]
Istvan Montvay\\
\textit{\large Deutsches Elektronen-Synchrotron DESY}\\
\textit{\large Notkestr.~85, D-22607 Hamburg, Germany}\\
\textit{\large E-mail: \email{montvay@mail.desy.de}}\\[5mm]
Stefano Piemonte\\
\textit{\large University of Regensburg, Institute for Theoretical Physics}\\
\textit{\large Universit\"atsstr.~31, D-93040 Regensburg, Germany}\\
\textit{\large E-mail: \email{stefano.piemonte@ur.de}}
\vspace*{5mm}}

\date{December 12, 2017}
\begin{document}
\maketitle

\newpage

\begin{abstract}
\noindent
\textbf{\textsf{Abstract:}}
In this work we present the results of a numerical investigation of SU(2)
gauge theory with $N_f = 3/2$ flavours of fermions, corresponding to 3
Majorana fermions, which transform in the adjoint representation of the
gauge group. At two values of the gauge coupling, the masses of bound states
are considered as a function of the fundamental fermion mass, represented by
the PCAC quark mass. The scaling of bound states masses indicates an
infrared conformal behaviour of the theory. We obtain estimates for the
fixed-point value of the mass anomalous dimension $\gamma^{\ast}$ from the
scaling of masses and from the scaling of the mode number of the
Wilson-Dirac operator. The difference of the estimates at the two gauge
couplings should be due to scaling violations and lattice spacing effects.
The more reliable estimate at the smaller gauge coupling is $\gamma^{\ast}
\approx 0.38(2)$.
\end{abstract}

\section{Introduction}

The long and short distance behaviour of QCD-like theories depends
significantly on the number $N_f$ of fermion flavours and on the
representation of the gauge group under which the fermions transform. For
sufficiently small $N_f$ the $\beta$-function is negative and the well-known
scenario with confinement and asymptotic freedom occurs. However, for large
$N_f$ above a certain limit $N_f^u$ asymptotic freedom is lost, the
$\beta$-function is positive and has an infrared fixed point at the origin.
The theory then has a scaling behaviour like $\phi^4$-theory and is
non-interacting in the continuum limit. Between these two cases different
interesting scenarios are expected. For $N_f$ below the triviality limit
$N_f^u$, the perturbative $\beta$-function to two or three loops shows
asymptotic freedom near the origin, but develops an infrared fixed point,
called Banks-Zaks fixed point, at a finite value of the coupling
\cite{Banks:1981nn}. If $N_f$ is just below $N_f^u$, this fixed point is at
weak couplings, and the scaling behaviour can be obtained perturbatively.
The theory is asymptotically free at short distances, and shows conformal
behaviour at large distances, i.\,e.\ it is infrared conformal. For smaller
$N_f$ the IR-fixed point moves towards stronger couplings, such that
perturbation theory ceases to be reliable. Finally, decreasing $N_f$ further
below a certain value $N_f^l$, the IR-fixed point will disappear and the QCD
scenario sets in. The region between $N_f^u$ and $N_f^l$ is the
\textit{conformal window}. Its upper edge can be estimated perturbatively,
but the determination of its lower edge is a non-perturbative problem.

\begin{figure}[ht]
\centering
\includegraphics[width=0.5\textwidth]{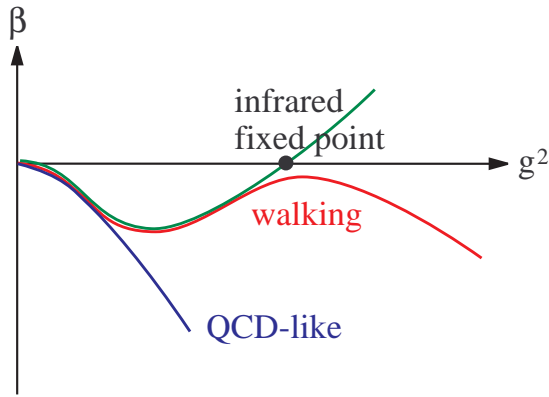}
\caption{Sketch of the $\beta$-functions for QCD-like, conformal, and
walking scenarios.}
\label{fig:beta-function}
\end{figure}

For a theory with $N_f$ below, but near to $N_f^l$, the $\beta$-function is
always negative, but might approach zero near a certain finite value of the
coupling, see Fig.~\ref{fig:beta-function}. In this case the coupling will
not run, but evolve rather slowly in a certain range of distances or
momenta, respectively. Such a behaviour is called \textit{walking}, and in
the walking regime the theory behaves approximately conformal
\cite{Holdom:1981rm}.

For a given theory, the question to which of these scenarios it belongs, is
of fundamental importance for its characteristic features. Based on the
perturbative $\beta$-function and approximate solutions of the
Schwinger-Dyson equations, Dietrich and Sannino \cite{Dietrich:2006cm} have
mapped out the phase diagram for non-supersymmetric theories with fermions
in different representations of the gauge group SU($N$) as a function of $N$
and $N_f$. It turns out that for representations higher than the fundamental
one, the boundaries of the conformal window are expected to be at rather
small values of $N_f$. In particular, for the adjoint representation they
might be near $N_f = 1$ and 2. To locate their true values requires,
however, non-perturbative methods. Therefore, in recent years the scaling
behaviour of various theories has been investigated by means of Monte Carlo
simulations in the framework of lattice gauge theory.

SU($N$) gauge theory with fermions in the adjoint representation has been
studied for $N_f = 2$ by various collaborations and appears to be
IR-conformal, see e.\,g.\ \cite{Bergner:2016hip} and
\cite{DeGrand:2015zxa,Pica:2017gcb} for reviews. A study of SU(2) with $N_f
= 1$ fermion flavours \cite{Athenodorou:2014eua} gives indications for
IR-conformal behaviour, too. The case of $N_f = 1/2$ describes one flavour
of Majorana fermions and corresponds to $\mathcal{N} = 1$ supersymmetric
Yang-Mills theory, which has been studied by our collaboration, see
\cite{Bergner:2015adz} and references therein. This theory is QCD-like
concerning its scaling behaviour.

It is the purpose of this article to present results about SU(2) gauge
theory with $N_f = 3/2$ flavours of fermions in the adjoint representation
of the gauge group, where 3/2 means 3 flavours of Majorana (or Weyl)
fermions. Preliminary results have been presented in \cite{Bergner:2017bky}.
We have investigated the masses of various particles, including mesons,
glueballs and spin 1/2 fermion-glue bound states, the string tension, and
the mass anomalous dimension, in order to gain insights into the IR
behaviour of the theory.

\section{Gauge theory with adjoint fermions on the lattice}

We consider SU(2) gauge theory coupled to fermions transforming under the
adjoint representation of the gauge group. In the continuum the covariant
derivative acting on a fermion field
\begin{equation}
\psi(x) = \psi^{a}(x) T^{a}\,,
\end{equation}
where $T^{a} = \sigma^{a} / 2$, $a=1,2,3$, are the generators of SU(2),
is given by
\begin{equation}
\mathcal{D}_{\mu} \psi(x) 
= (\partial_{\mu} \psi(x) + \I\, g [A_{\mu}(x), \psi(x)]),
\end{equation}
with the gauge field $A_{\mu}(x) = A_{\mu}^{a}(x)\, T^{a}$.

The lattice formulation of the theory that we use employs the Wilson gauge
action built from the plaquette variables $U_p$ and the Wilson-Dirac
operator in the adjoint representation. The lattice action is
\begin{equation}
\mathcal{S}_L =
\beta \sum_p\left(1-\frac{1}{2} \mathrm{tr}\, U_p\right)
+\sum_{xy,f} \bar{\psi}_x^{f}(D_w)_{xy}\psi_y^{f}\,,
\end{equation}
where $D_w$ is the Wilson-Dirac operator
\begin{multline}
(D_w)_{x,a,\alpha;y,b,\beta}
    =\delta_{xy}\delta_{a,b}\delta_{\alpha,\beta}\\
    -\kappa\sum_{\mu=1}^{4}
      \left[(1-\gamma_\mu)_{\alpha,\beta}(V_\mu(x))_{ab}
                          \delta_{x+\mu,y}
      +(1+\gamma_\mu)_{\alpha,\beta}(V^\dag_\mu(x-\mu))_{ab}
                          \delta_{x-\mu,y}\right].
\end{multline}
Here $\beta=1/g^2$ is the inverse bare gauge coupling, and the hopping
parameter $\kappa$ is related to the bare fermion mass via
$\kappa=1/(2m_0+8)$. The link variables $U_{\mu}(x)$ are in the fundamental
representation of the gauge group SU(2). The gauge field variables
$V_{\mu}(x)$ in the adjoint representation are given by
$[V_\mu(x)]^{ab}=2\,\mathrm{tr} [U^\dag_\mu(x) T^a U_\mu(x) T^b]$.

The lattice extension in all spatial directions is denoted by the number
$N_s$ of lattice points. In our simulations the extension in the temporal
direction is always given by $N_t = 2 N_s$.

The number of fermion flavours is conventionally counted in terms of Dirac
fermions. Majorana fermions, satisfying the Majorana condition
\begin{equation}
\overline{\psi} = \psi^{T} C,
\end{equation}
where $C$ is the charge conjugation matrix, possess half the number of
degrees of freedom as Dirac fermions and are counted as $N_f = 1/2$.
Consequently $N_f = 3/2$ is to be interpreted as 3 species of Majorana
fermions. In this case the index $f$ in the lattice action counts Majorana
fermions and runs from 1 to 3.

In order to reduce lattice artefacts we use in our simulations an improved
version of the lattice action with a tree-level Symanzik improved gauge
action and stout smearing for the link fields in the Wilson-Dirac operator
\cite{Morningstar:2003gk}. The stout smearing is iterated three times with
the smearing parameter $\rho=0.12$.

For Majorana fermions the fermion integration
\begin{equation}
\int [d\psi]\ \E^{- \frac{1}{2} \overline{\psi} D_w \psi}
= \mathrm{Pf}(C D_w) = \pm \sqrt{\det D_w}
\end{equation}
yields the Pfaffian of the Wilson-Dirac matrix. With 3 Majorana fermion
fields the functional integrals contain a factor $(\det D_w)^{3/2}$, which
can be treated with the PHMC algorithm. The possible sign of Pf($C D_w$) has
to be taken into account in the observables by reweighting. In simulations
not too close to the critical hopping parameter $\kappa_c$, negative signs
are very rare and it was not necessary to consider them in the parameter
regions of our simulations for the determination of the masses.

\begin{figure}[ht]
\centering
\includegraphics[width=0.8\textwidth]{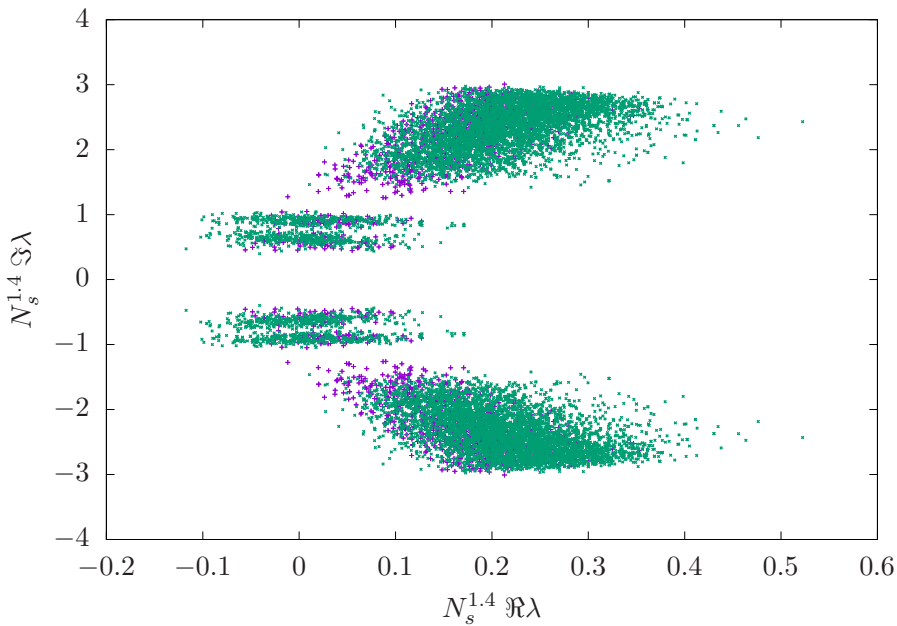}
\caption{The eigenvalues of the Wilson-Dirac operator for the near-critical
runs R1 ($N_s = 24$) and R3 ($N_s = 32$), represented in the complex plane.
The smallest eigenvalues in a region around the real axis are determined. We
observe a good agreement between the two runs by rescaling the eigenvalues
with $N_s^{1.4}$. This scaling of the eigenvalues is in accordance with the
scaling by $1 + \gamma^{\ast}$ investigated in \cite{Karthik:2016ppr},
assuming $\gamma^{\ast} \approx 0.4$.}
\label{fig:eigev}
\end{figure}

In order to check the possible presence of a negative sign we have generated
two runs at the critical parameters corresponding to $\mpcac \approx 0$
(runs R1 and R3, see Table \ref{tab:parameters}). The eigenvalue
distribution for these runs does not completely match the bounds of the
polynomial approximation, but they can still be used to check the general
properties of the sign problem in this theory without determining the
otherwise necessary correction factors on the configurations. We observe
that even at these critical parameters no negative sign is obtained for the
measured configurations and a gap in the imaginary part Wilson-Dirac
eigenvalues around zero appears. The general features of the spectrum scale
with the volume (see Fig.~\ref{fig:eigev}). Consequently the rather large
finite size effects in these runs are most likely preventing a fluctuating
Pfaffian sign in this theory. We did not observe such an effect in
supersymmetric Yang-Mills theory.

For generating field configurations on the lattice we have used the two-step
polynomial hybrid Monte Carlo (PHMC) algorithm \cite{Montvay:2005tj}. It is
based on polynomial approximations of the inverse powers of the Wilson-Dirac
matrix. The first polynomial gives a rough approximation that is corrected
by the second polynomial. The polynomials were chosen such that the lower
bound of the approximation interval was about a factor 10 smaller than the
smallest occurring eigenvalues. The resulting two-step approximation is so
good that no further correction by a reweighting factor is necessary in
practice.

\section{Model parameters and continuum limit}

In an asymptotically free gauge theory the lattice spacing $a$ mainly
depends on the gauge coupling $\beta$. For increasing $\beta$ the lattice
spacing decreases exponentially. The size of $a$ in physical units can be
fixed in terms of a dimensionful quantity like the Sommer scale $r_0$ or the
string tension $\sigma$, which is used to set the scale.

In an IR conformal theory the situation is different. In the close vicinity
of the fixed point the coupling $\beta$ is an irrelevant parameter and the
model depends only weakly on it. Due to the absence of a mass scale the size
of the lattice spacing can only be compared to the physical extent $L = N_s
a$ of the lattice, and the continuum limit has to be defined in terms of the
ratio $a/L$.

Nevertheless, away from the IR fixed point towards the Gaussian fixed point
at $g^2 = 0$ a relevant dependence on $\beta$ is expected. The theory is
asymptotically free in the ultraviolet, and the continuum limit would be
reached by sending $\beta$ to infinity. Near the IR fixed point the
dependence on $\beta$ appears as a correction to scaling.

In addition, the mass term in the action plays an important role. Non-zero
masses are relevant parameters that break conformal symmetry and imply
corrections to scaling. In the presence of mass terms the renormalisation
flow doesn't run into the IR fixed point, but may pass close by. The running
of $\beta$ is then expected to be rather slow.

\begin{figure}[ht]
\centering
\includegraphics[width=0.49\textwidth]{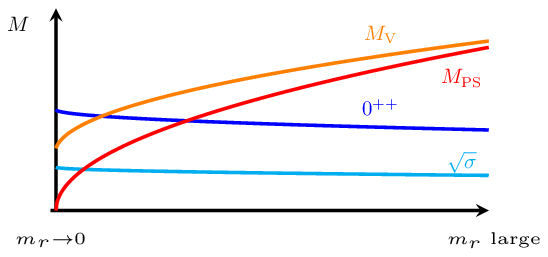}
\includegraphics[width=0.49\textwidth]{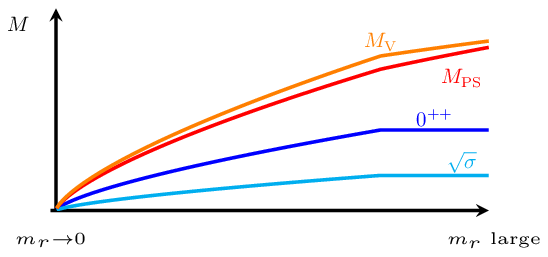}
\caption{Sketch of the mass spectrum as a function of the fermion mass $m_r$
for a QCD-like (left) and a IR conformal (right) scenario. Indicated are the
masses of the pseudoscalar (PS) and vector (V) mesons, the scalar (0++)
glueball and the square-root of the string tension $\sigma$.}
\label{fig:spectrum}
\end{figure}

The dependence of particle masses on the renormalised fermion mass $m_r$
would be quite different for theories in the different scenarios. In a
theory above the conformal window, with confinement and chiral symmetry
breaking, the mass of the pseudo-Goldstone boson vanishes when the fermion
mass $m_r$ goes to zero, whereas the other particle masses approach a finite
value, see Fig.~\ref{fig:spectrum}.

In the IR conformal scenario all particle masses and the
string tension would asymptotically scale to zero in the conformal limit
according to
\begin{equation}
M \propto m_r^{1/(1 + \gamma^{\ast})}\,,
\end{equation}
where $\gamma^{\ast}$ is the value of the mass anomalous dimension at the
fixed point \cite{Luty:2008vs,DelDebbio:2010ze}.

In this scenario the ratios of masses are approximately constant for small
$m_r$. These ratios represent universal features of (near) IR conformal
theories \cite{Athenodorou:2016ndx}.

In practice, however, the limit of vanishing fermion mass $m_r$ cannot be
reached in numerical simulations. In a near conformal theory severe finite
size effects would occur for small $m_r$, and have a substantial influence
on the mass spectrum. Moreover, the simulation algorithms slow down strongly
at small fermion masses and in this way limit the accessible parameter
range.

In our simulations we have chosen two values of $\beta = 1.5$ and $1.7$. We
checked that these inverse gauge couplings are above the value of the bulk
(``finite temperature'') phase transition. In order to control finite volume
effects, lattices of size $16^3 \times 32$, $24^3 \times 48$ and $32^3
\times 64$ have been used for both values of $\beta$.

The renormalised fermion mass has been varied by a series of values for the
hopping parameter $\kappa$. As renormalised fermion mass $m_r$ we take the
PCAC mass $\mpcac$, determined by means of the PCAC (partially conserved
axial current) relation. The lattice sizes and parameters for the ensembles
with positive fermion mass are summarised in Tab.~\ref{tab:parameters}.

\begin{table}[ht!]
\begin{center}
\begin{tabular}{|c|c|c|l|c|}
\hline
 & $\beta$ & $N_s$ & \hspace{10pt}$\kappa$ & $a \mpcac$ \\
\hline
A & 1.5 & 16 & 0.137  &  0.02270(18) \\
B & 1.5 & 16 & 0.135  &  0.11604(44) \\
C & 1.5 & 16 & 0.132  &  0.23236(83) \\
D & 1.5 & 24 & 0.1351 &  0.10986(12) \\
E & 1.5 & 24 & 0.134  &  0.15632(15) \\
F & 1.5 & 24 & 0.133  &  0.19515(20) \\
G & 1.5 & 24 & 0.132  &  0.23207(22) \\
H & 1.5 & 32 & 0.1359 &  0.07380(07) \\
\hline
J & 1.7 & 16 & 0.130  &  0.12890(77) \\
K & 1.7 & 24 & 0.133  &  0.03360(30) \\
L & 1.7 & 24 & 0.132  &  0.06628(08) \\
M & 1.7 & 24 & 0.130  &  0.12882(15) \\
N & 1.7 & 32 & 0.132  &  0.06635(12) \\
O & 1.7 & 32 & 0.130  &  0.12910(04) \\
P & 1.7 & 32 & 0.1285 &  0.17366(04) \\
Q & 1.7 & 48 & 0.1322 &  0.05990(05) \\
R1& 1.7 & 24 & 0.134  & -0.00097(22) \\
R3& 1.7 & 32 & 0.134  & -0.00052(11) \\
\hline
\end{tabular}
\caption{Parameters of the simulation ensembles.}
\label{tab:parameters}
\end{center}
\end{table}

The mass of the pseudoscalar meson in lattice units was in the range 0.13 to
1.0. Most relevant for finite size effects are the lightest particle masses,
which in our simulations turned out to be the scalar glueball and the
pseudoscalar meson. As the mass of the pseudoscalar meson can be determined
much easier and precisely, we consider this one as a measure for the low
mass scale.

Our results for the various masses show that at $\beta = 1.5$ ensemble A,
and at $\beta = 1.7$ ensembles J and K have sizeable finite size effects, so
that these ensembles are usually discarded in the analysis, apart from the
cases where finite size scaling effects are included.

\section{Scaling of the lightest particle masses}

The spectrum of colour neutral particles in this model consists of bound
states of gauge bosons (``gluons'') and fermions. In addition to mesonic
particles and glueballs, spin 1/2 bound states of gluons and fermions are
possible due to the adjoint representation of the fermions. In the mesonic
sector we consider the scalar and pseudoscalar ones, created by the
operators $\bar{\psi} \sigma^{a} \psi$ and $\bar{\psi} \gamma_5 \sigma^{a}
\psi$, $a=1,2,3$, respectively, and the vector and pseudovector mesons,
created by $\bar{\psi}^b \gamma_k \psi^c$ and $\bar{\psi}^b \gamma_5
\gamma_k \psi^c$, respectively, where $k=1,2,3$ is in the spatial direction.
In addition, the scalar glueball and the spin 1/2 fermion-glue bound state,
represented by $\sigma_{\mu\nu} \mathrm{tr}\left[F^{\mu\nu} \psi \right]$,
have been investigated. Apart from the particle masses we have also
calculated the string tension $\sigma$ from the static quark-antiquark
potential, where ``quark'' means a particle in the fundamental
representation of the colour gauge group. The square-root of $\sigma$ has
dimensions of a mass and scales as a mass. Therefore we include it in our
analysis of the scaling behaviour. The techniques for the calculation of the
propagators, masses and string tension have been explicated in
\cite{Bergner:2016hip} and for details we refer to this article.

\begin{figure}[ht!]
\centering
\includegraphics[width=0.8\textwidth]{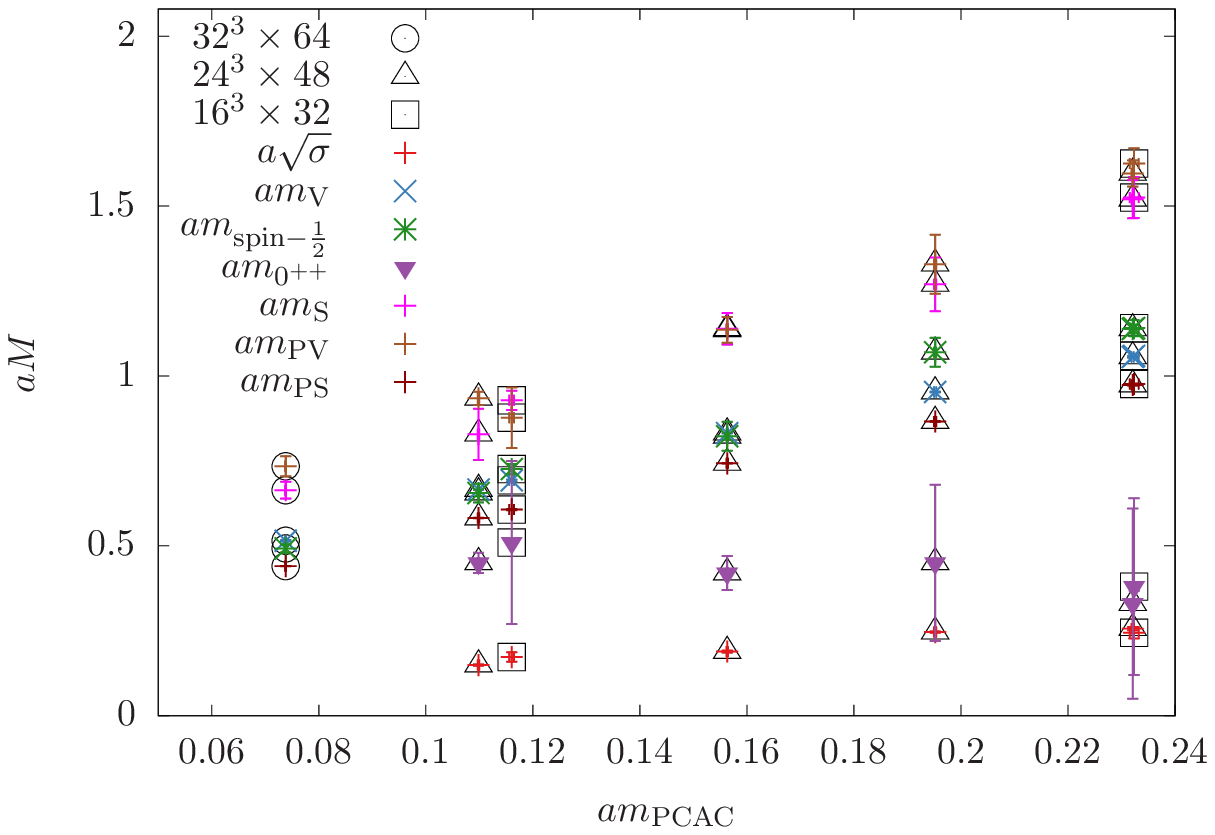}\\
\includegraphics[width=0.8\textwidth]{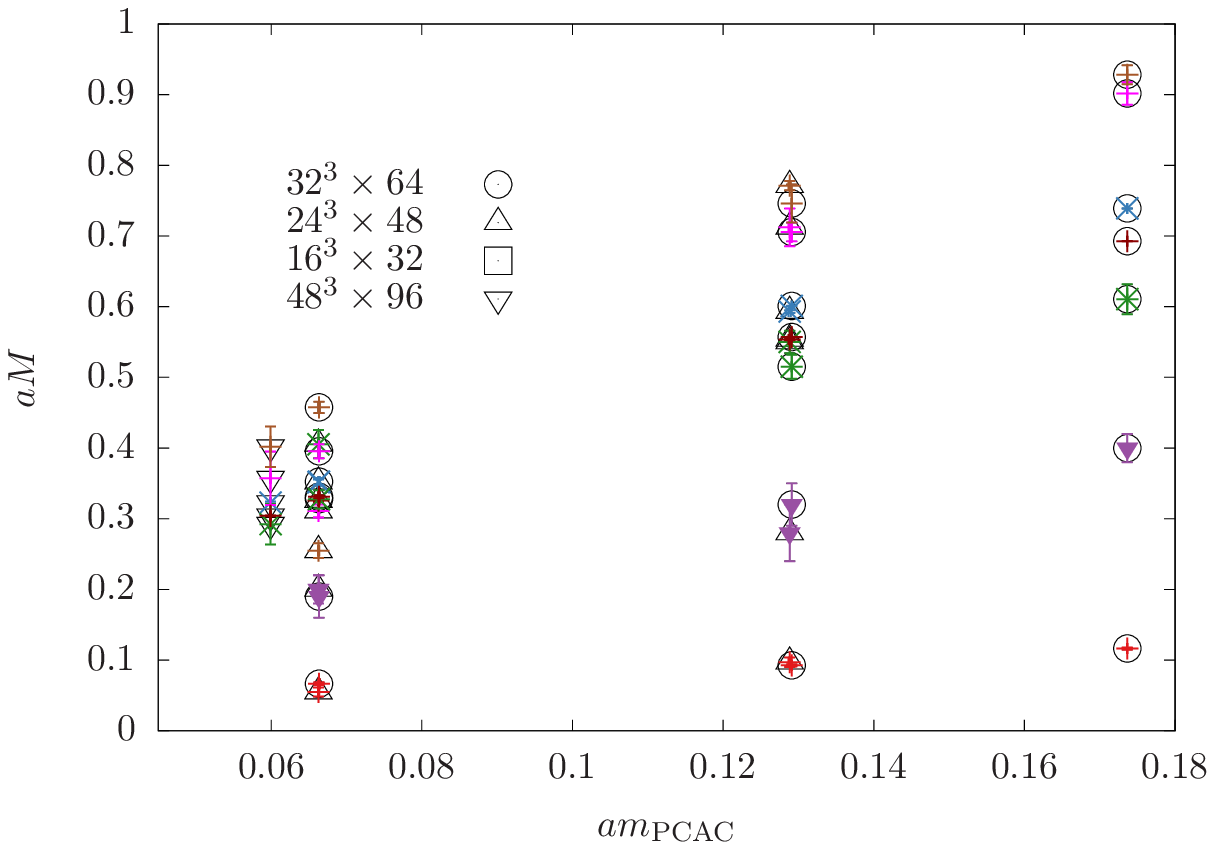}
\caption{Particle masses and $\sqrt{\sigma}$ as a function of $a \mpcac$
for $\beta=1.5$ (above) and $\beta=1.7$ (below).}
\label{fig:masses}
\end{figure}
\begin{figure}[ht!]
\centering
\subfigure{\includegraphics[width=0.49\textwidth]{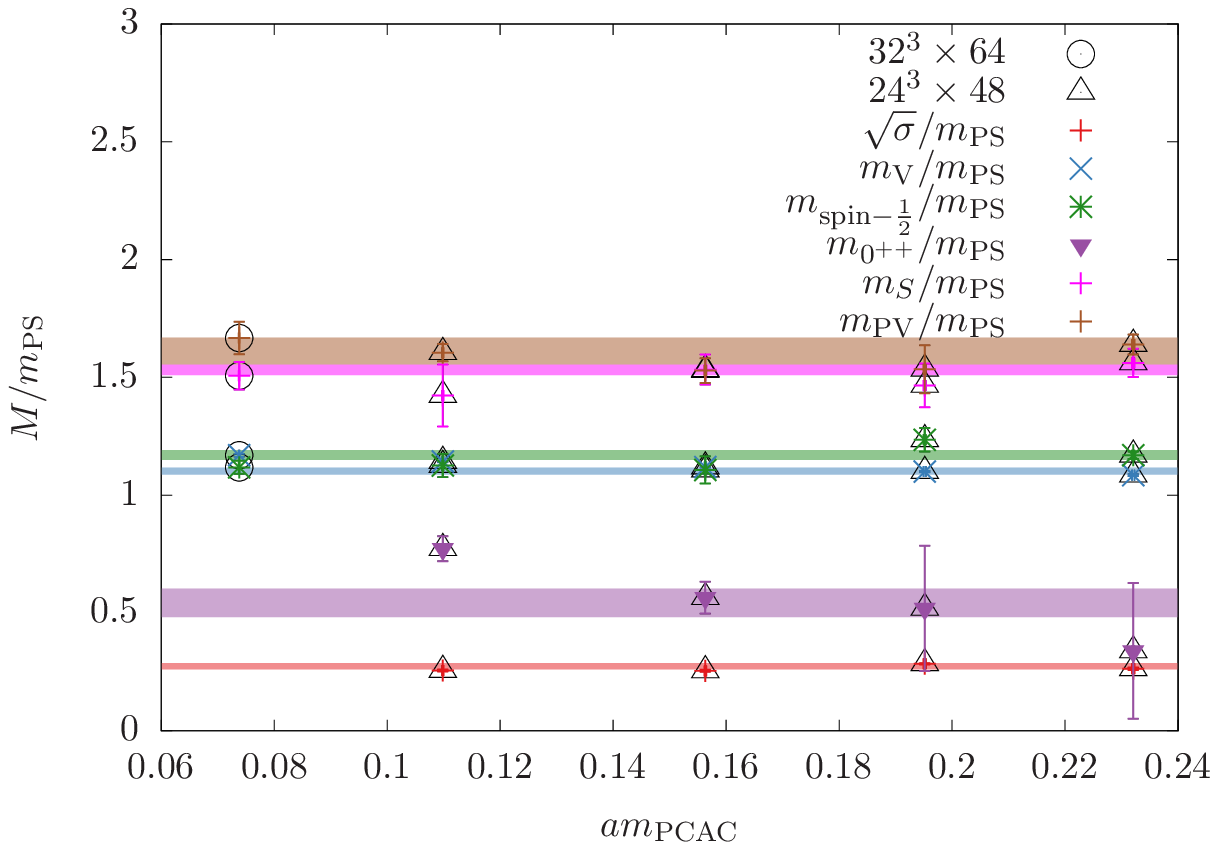}}
\subfigure{\includegraphics[width=0.49\textwidth]{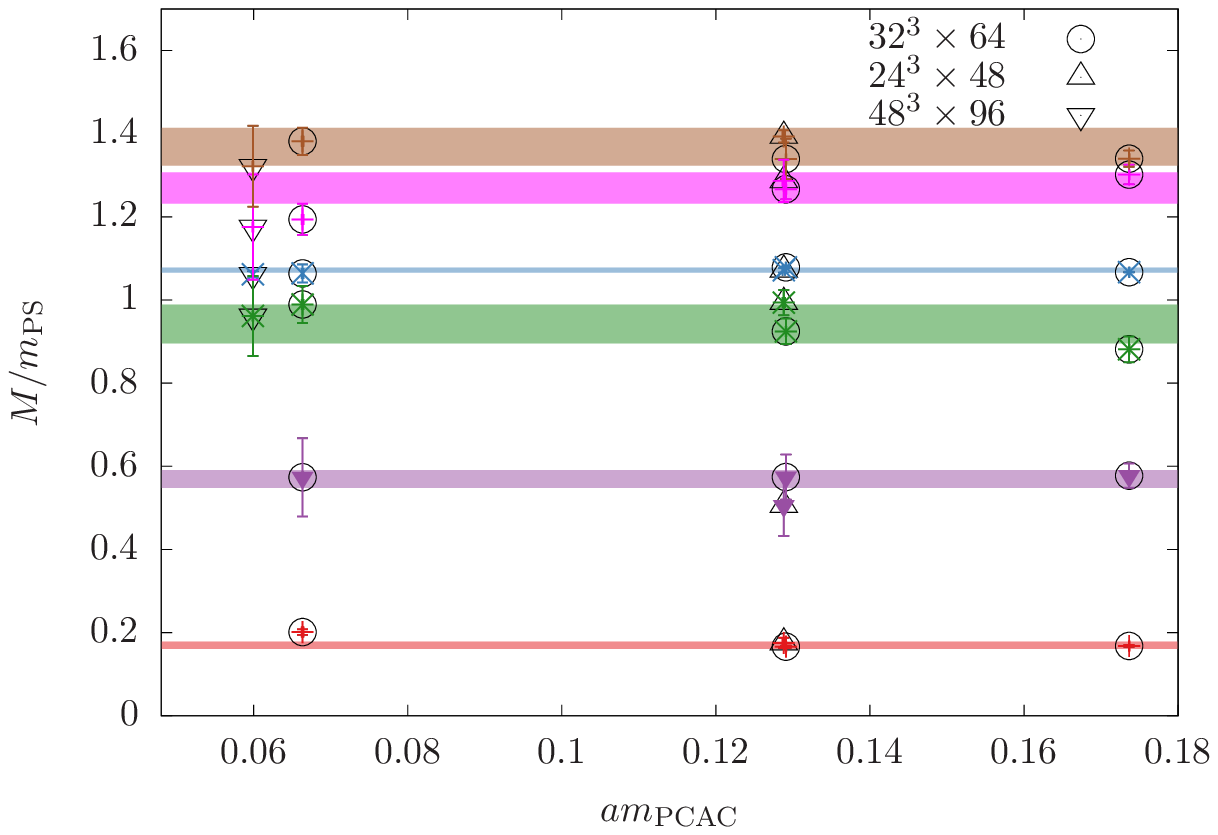}}
\caption{Particle masses and $\sqrt{\sigma}$ in units of the pseudoscalar
mass as a function of $a \mpcac$ for $\beta=1.5$ and $\beta=1.7$.}
\label{fig:ratios}
\end{figure}

Fig.~\ref{fig:masses} shows the particle masses as a function of the fermion
mass. All masses appear to scale downwards towards the limit $\mpcac = 0$.
The lightest particle, being well separated from the rest, is the scalar
glueball. So the overall behaviour indicates a scenario different from the
QCD-like one, where the pseudoscalar pseudo-Goldstone boson is lightest
particle. As expected for a theory in the conformal window, all masses scale
approximately in the same way and their ratios are constant as shown in
Fig.~\ref{fig:ratios}.

In order to substantiate this impression we have investigated the scaling
behaviour of masses. To begin with, consider the pseudoscalar meson. In a
QCD-like situation this particle is the pseudo-Goldstone boson of
spontaneously broken chiral symmetry, and its mass vanishes with $m_r \equiv
\mpcac$ according to the Gell-Mann-Oakes-Renner relation $\mps \propto
m_r^{1/2}$. On the other hand, in an IR-conformal scenario $\mps$ would
scale to zero as $\mps \propto m_r^k$ with an exponent $k = 1/(1 +
\gamma^{\ast})$ that can be different from 1/2. In order to determine the
exponent $k$ we fitted $\ln (\mps)$ as a linear function of $\ln (m_r)$. As
mentioned above, ensembles A, K and J are omitted in view of finite size
effects. Being even more restrictive concerning possible finite size
effects, one would also leave out the ensembles N, L and Q with the smallest
fermion masses at $\beta = 1.7$, remaining with M, O, P.

The data points show a nice linear behaviour. For $\beta = 1.5$ the fit
gives an exponent $k = 0.691(2)$, and for $\beta = 1.7$ we obtain $k =
0.743(14)$ from ensembles (M, O, P). An estimate of the systematic error is
obtained by considering different subsets of ensembles. For $\beta = 1.7$ we
get $k = 0.775(8)$ from the $N_s = 32, 48$ lattices (N, O, P, Q), and $k =
0.780(7)$ from ensembles L -- Q. Thus the exponent is evidently different
from 0.5, indicating the IR-conformal scenario. In this case, the other
masses should show scaling with the same exponent. We considered the
weighted average of the logarithms of the other masses $\ms, \mv, \mpv,
\mgb, \msh$ and $m_{\sigma} = \sqrt{\sigma}$, the weights given by the
inverse variances as usual, as a function of $\ln (m_r)$. Again a nice
linear behaviour can be seen. From the linear fit we obtain $k = 0.608(17)$
for $\beta = 1.5$ and $k = 0.667(54)$ for $\beta = 1.7$, ensembles M, O, P.
These values are compatible with the ones from $\mps$ and give clear support
for the IR-conformal scenario.

\begin{figure}[ht!]
\centering
\includegraphics[width=0.8\textwidth]{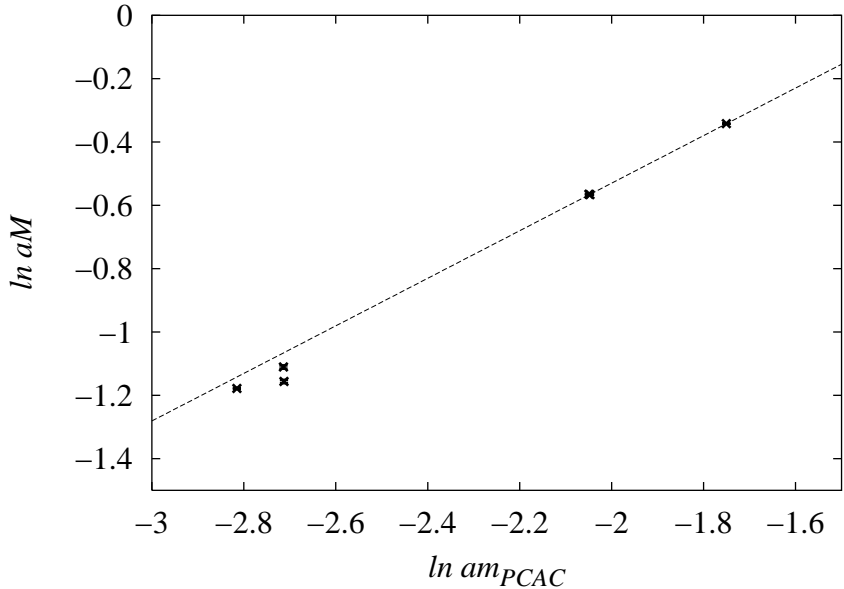}
\caption{Weighted averages of the logarithms of the particle masses at
$\beta = 1.7$ as a function of $\ln (a \mpcac)$, and the scaling fit with
exponent $k = 0.751$. Note that the second symbol from right stands for two
close-by data points and the three leftmost points are not included in this
fit.}
\label{fig:massfit}
\end{figure}

The overall estimate of the mass anomalous dimension is obtained by the same
fit, now including the pseudoscalar mass. For $\beta = 1.5$ this gives $k =
0.675(24)$. The corresponding number for $\beta = 1.7$, ensembles M, O, P, is
$k = 0.751(8)$. For comparison, we get $k = 0.817(36)$ from N, O, P, Q, and
$k = 0.817(24)$ from L -- P. Fig.~\ref{fig:massfit} shows the averaged
logarithmic masses for $\beta = 1.7$ and the fit with $k = 0.751$.

To conclude, the particle masses show scaling behaviour of the IR-conformal
scenario with an exponent $k \approx 0.67$, corresponding to $\gamma^{\ast}
\approx 0.5$, for $\beta = 1.5$, and $k = 0.75(7)$, corresponding to
$\gamma^{\ast} = 0.33(13)$, for $\beta = 1.7$.

\section{Mode number}

An alternative method for the determination of the mass anomalous dimension
is based on the spectral density of the Dirac operator
\cite{Giusti:2008vb,Patella:2012da,Cheng:2013bca,Fodor:2014zca,Cichy:2013eoa}.
The mode number $\nu(\Omega)$ is defined to be the number of eigenvalues of
the hermitian operator $D_w^\dag D_w$ below some limit $\Omega^2$. The mode
number obeys a scaling law \cite{Patella:2012da}
\begin{equation}
\label{eq:modenumber}
\nu(\Omega) = \nu_0 + a_1 (\Omega^2 - a_2^2)^{2/(1 + \gamma^{\ast})}
\end{equation}
for sufficiently small values of $\Omega^2 - a_2^2$, where $a_2$ is
proportional to $\mpcac$. Therefore, a fit of $\nu(\Omega)$ to this function
in a suitable range $[\Omega_{\text{min}},\Omega_{\text{max}}]$ allows to
estimate the mass anomalous dimension $\gamma^{\ast}$.

The choice of the fit range $[\Omega_{\text{min}},\Omega_{\text{max}}]$ is a
sensitive issue. For a small fit range near a scale $\Omega$, the resulting
value for the mass anomalous dimension can be considered as an effective
anomalous dimension $\gamma(\Omega)$, which approximates the corresponding
renormalisation group function \cite{Cheng:2013bca}. For large $\Omega$ it
is expected that $\gamma(\Omega)$ decreases and approaches its value zero at
the asymptotically free UV fixed point. On the other hand, for small
$\Omega$ finite volume effects and effects of the non-vanishing fermion mass
$\mpcac$ will disturb the scaling behaviour. Therefore the fit range should
be located in an intermediate regime, where the effects of the finite volume
and non-zero fermion mass can be neglected
\cite{Patella:2012da,Cichy:2013eoa}. For an infrared conformal theory the
coupling runs very slowly for a wide range of scales at low $\mu$, and there
the anomalous dimension $\gamma$ varies slowly, too, approximatively
developing a plateau at the value $\gamma^{\ast}$. Investigations of the
$\beta$-function in the MiniMOM scheme for this theory
\cite{Bergner:2017ytp} indicate that the $N_f=3/2$ theory appears to be
close to the the edge of the conformal window.

\begin{figure}
\centering
\includegraphics[width=0.6\textwidth]{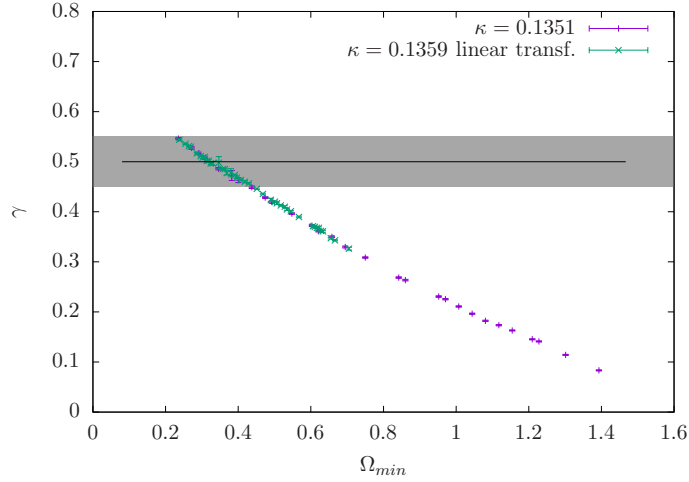}
\caption{Fitted value of the mass anomalous dimension for the ensemble at
$\beta=1.5$, $\kappa=0.1351$ (ensemble D), and $\kappa=0.1359$ (ensemble H).
$\Omega_{\text{min}}$ denotes the lower end of the fitting interval, while
the upper end is fixed to $\Omega_{\text{min}} + 0.07$. The
$\Omega_{\text{min}}$ values on the x-axis are rescaled by the linear
transformation $\Omega_{\text{min}}'=0.88 \Omega_{\text{min}} + 0.1$ for
$\kappa=0.1359$ in order to collapse the two ensembles. In the shaded region
we obtain the best fits for both ensembles. Fits with a correlated $\chi^2$
per degree of freedom larger than 4 are excluded.}
\label{mode150}
\end{figure}

\begin{figure}
\centering
\includegraphics[width=0.6\textwidth]{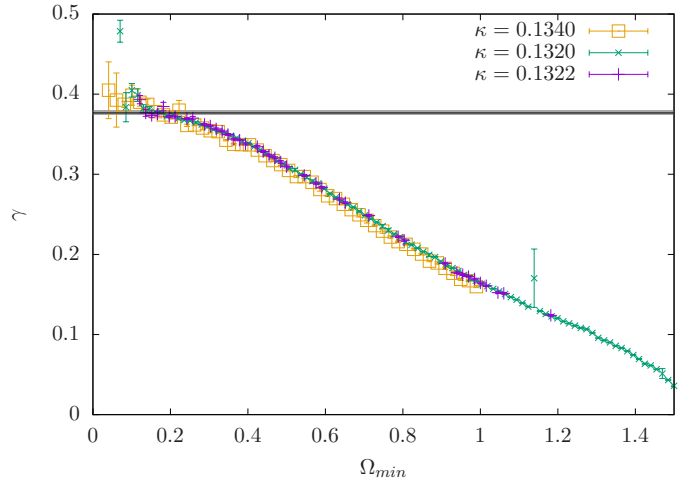}
\caption{Fitted value of the mass anomalous dimension for the ensemble at
$\beta=1.7$, $\kappa=0.1322$ (ensemble Q), $\kappa=0.1320$ (ensemble N), and
$\kappa=0.1340$ (ensemble R3), as in Figure~\ref{mode150}. Ensembles N and
R3 are only shown for comparison. Fits of the data from ensemble Q with a
correlated $\chi^2$ per degree of freedom larger than 4 have been excluded
from the figure. The final value represented by the shaded region in the
plot is only obtained from a fit in the plateau region of ensemble Q.}
\label{mode170}
\end{figure}

The techniques and the code that we have implemented to compute the mode
number have been tested in the $N_f=2$ case and presented in
Ref.~\cite{Bergner:2017bky}. We have used different methods for the
non-linear fit to Eq.~\ref{eq:modenumber} including parallel tempering
together with conjugate gradient techniques. The fit of a large number of
data points requires particular techniques for the determination of the
correlated $\chi^2$ \cite{Michael:1994sz}. The results for the fits over
intervals of fixed length and varying lower end $\Omega_{\text{min}}$ are
shown in Figs.~\ref{mode150} and \ref{mode170} for the runs at $\beta=1.5$
and $\beta=1.7$, respectively.


At $\beta=1.5$ we obtain reasonable fits with an acceptable p-value and a
correlated $\chi^2$ per degree of freedom in a certain region of $\Omega$
values for the ensembles with the smallest fermion masses. However, there is
no pronounced plateau for the obtained values in this range. The best fits
are obtained at rather large values of $\Omega$. Further in the infrared,
the correlated $\chi^2$ of the fit drastically increases, which is an
indication of fermion mass effects. We take the final value from the middle
of the range where the correlated $\chi^2$ per degree of freedom is below
2.5 for ensemble H, and the width of this range as an estimate for the
error. This provides a rough estimate of $\gamma^{\ast} \approx 0.5 \pm
0.05$.

In contrast to the case of $\beta=1.5$, a considerable plateau of the fitted
values is obtained at $\beta=1.7$ in the infrared region. The plots of the
fit results agree in a large range of $\Omega_{\text{min}}$ values for the
ensembles Q and N. Hence they are quite insensitive to a change of the
fermion mass and the volume. Even the plot for ensemble R3, at approximate
zero fermion mass, agrees with these data. Due to the uncertainties
originating from the polynomial choices in the PHMC algorithm at ensemble
R3, we have only considered ensemble Q for the final fit. We obtain a value
of $\gamma^{\ast} \approx 0.377(3)$. Taking also the uncertainties in the
determination of the fitting interval into account, the estimate is
$\gamma^{\ast} \approx 0.38(2)$.

\begin{figure}[ht!]
\centering
\subfigure{\includegraphics[width=0.49\textwidth]{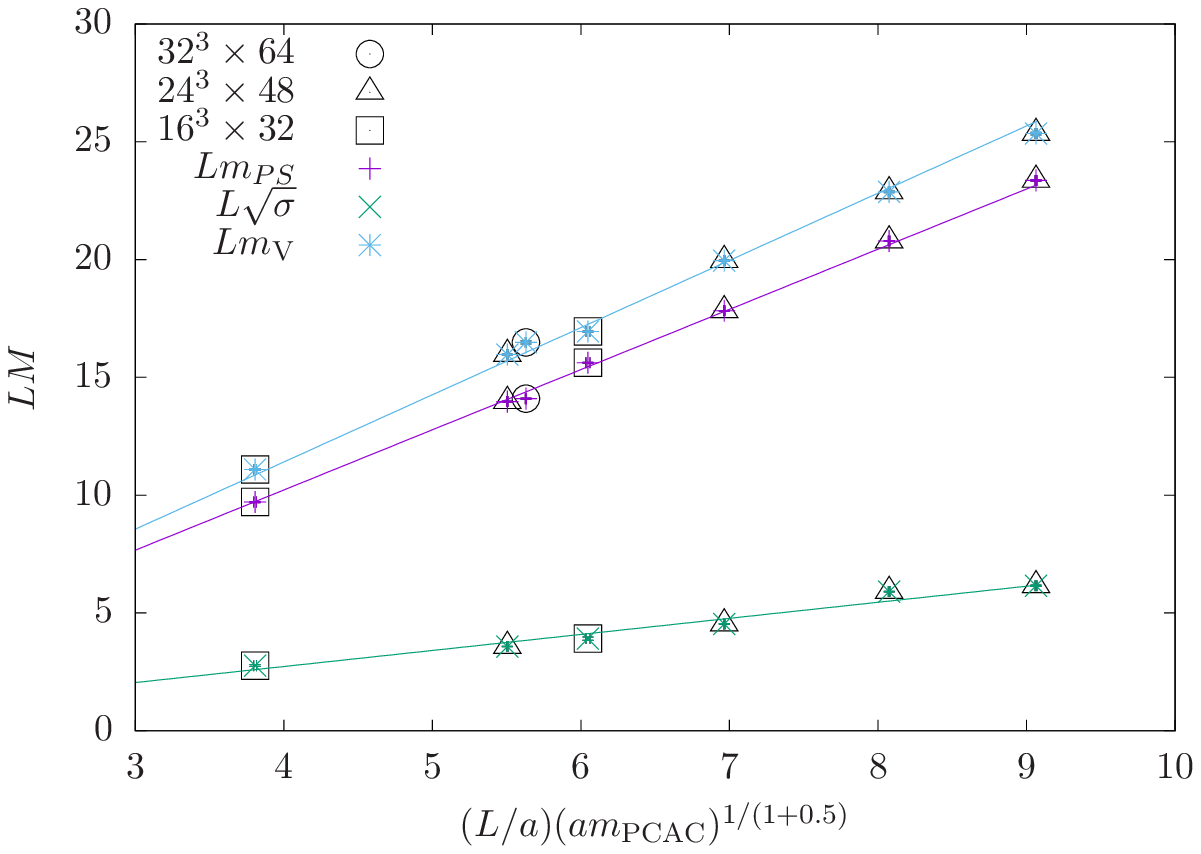}}
\subfigure{\includegraphics[width=0.49\textwidth]{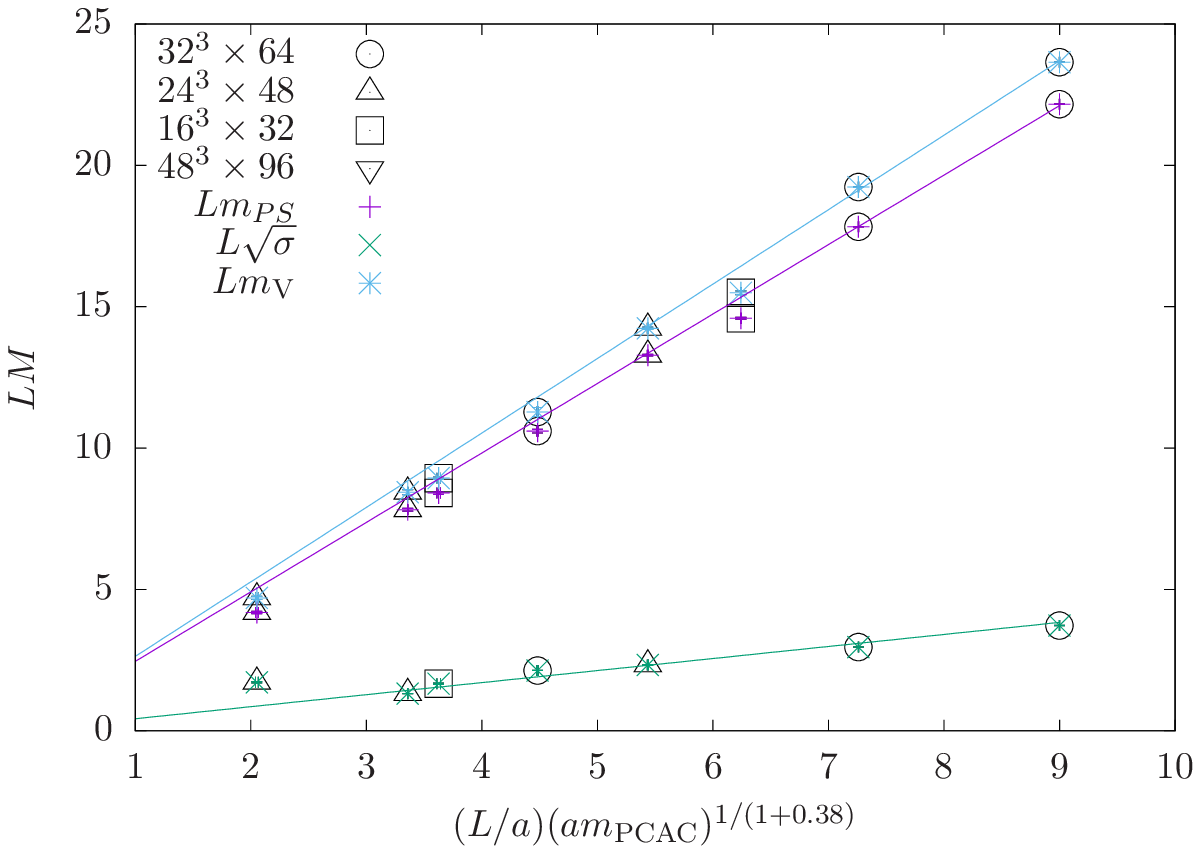}}
\caption{A cross check of the scaling exponents obtained from the mode
number with the scaling of the particle masses. These figures show a fit to
the hyperscaling hypothesis of the masses including volume scaling. The
points with the smallest values on the x-axis correspond to the ensembles B
($\beta=1.5$) and K ($\beta=1.7$). The lines correspond to a linear fit, in
case of $\beta=1.7$ without the data of ensemble K.}
\label{fig:scaling}
\end{figure}

We made a crosscheck of the obtained values of $\gamma^{\ast}$ with the
hyperscaling of the mass spectrum. As shown in Fig.~\ref{fig:scaling}, the
agreement with the expected functional behaviour is reasonable. We can also
vary the exponents close to the measured values in order to minimise the sum
of the $\chi^2$ from the linear fits. In this way we obtain a minimum around
$\gamma^{\ast} \approx 0.46(2)$ for $\beta=1.5$ and $\gamma^{\ast} \approx
0.37(2)$ for $\beta=1.7$. This shows that the values for the mass anomalous
dimension obtained from the mode number are consistent with the hyperscaling
of the mass spectrum

\section{Conclusions}

We have analysed the spectrum of bound states masses in SU(2) gauge theory
with $N_f = 3/2$ flavours of adjoint fermions at two values of the inverse
gauge coupling $\beta = 1.5$ and $1.7$. The scaling of the masses as a
function of the fermion mass $\mpcac$ indicates an infrared conformal
behaviour of this theory. The fixed point value of the mass anomalous
dimension is estimated to be $\gamma^{\ast} \approx 0.5$, for $\beta = 1.5$,
and $\gamma^{\ast} = 0.33(13)$, for $\beta = 1.7$. An independent estimate
has been obtained from the scaling of the mode number of the hermitian
Wilson-Dirac operator. For $\beta = 1.5$ we only get a rough estimate of
$\gamma^{\ast} \approx 0.5$, whereas for $\beta = 1.7$ a plateau shows up at
a value of $\gamma^{\ast} \approx 0.38(2)$.

For a conformally behaving theory in the infinite volume limit the value of
$\gamma^{\ast}$ should be independent of the gauge coupling. On the other
hand, for a theory in the vicinity of an IR fixed point, scaling violations
are present, which increase towards the UV regime. The fact that our
estimates at the two gauge couplings do not exactly coincide indicates the
influence of scaling violations and cutoff effects.

\section*{Acknowledgments}

The authors gratefully acknowledge the Gauss Centre for Supercomputing (GCS)
for providing computing time for a GCS Large-Scale Project on the GCS share
of the supercomputer JUQUEEN at J\"ulich Supercomputing Centre (JSC) and on
the supercomputer SuperMUC at Leibniz Computing Centre (LRZ). GCS is the
alliance of the three national supercomputing centres HLRS (Universit\"at
Stuttgart), JSC (Forschungszentrum J\"ulich), and LRZ (Bayerische Akademie
der Wissenschaften), funded by the German Federal Ministry of Education and
Research (BMBF) and the German State Ministries for Research of
Baden-W\"urttemberg (MWK), Bayern (StMWFK) and Nordrhein-Westfalen (MIWF).
Further computing time has been provided by the compute cluster PALMA of the
University of M\"unster.


\end{document}